\documentclass[12pt,oneside]{article}   

\usepackage{rotating}
\usepackage{amsmath}
\usepackage{xcolor}         
\usepackage{fancyhdr}		
\usepackage[backend=biber,style=ieee]{biblatex}
\usepackage{makecell} 

\usepackage{caption}
\usepackage[section]{placeins}   %

\usepackage{graphicx}

\usepackage{chngcntr} 
\newcounter{supfigure}

\newenvironment{supfigure}[1][]{
  \refstepcounter{supfigure}%
  \renewcommand{\thefigure}{S\arabic{supfigure}}
  \begin{figure}[#1]
}{
  \end{figure}
  \renewcommand{\thefigure}{\arabic{figure}}
}

\makeatletter \renewcommand\@biblabel[1]{$^{#1}$} \makeatother
\usepackage{marginnote}
\newcommand{\mfig}[1]{\marginpar{{\sf Fig~\ref{#1} }}}
\newcommand{\mtab}[1]{\marginpar{{\sf Table~\ref{#1} }}}

\newcommand{\cen}[1]{\begin{center} #1 \end{center}}

\usepackage[mathlines]{lineno}

\linenumbers
\nolinenumbers

\usepackage{subcaption}
\usepackage{hyperref}
\hypersetup{ colorlinks,
    citecolor=blue,
    filecolor=blue,
    linkcolor=blue,
    urlcolor=blue
}

\usepackage{xcolor}

\definecolor{gray}{rgb}{0.6,0.6,0.6}
\definecolor{red}{rgb}{0.85,0,0}
\definecolor{green}{rgb}{0,0.85,0}
\definecolor{blue}{rgb}{0,0,0.85}
\definecolor{beige}{rgb}{0.92,0.87,0.78}
\usepackage[all]{hypcap}    

\usepackage{multirow}
\usepackage{array}

\begin{document}

\cen{
\sf {
\Large 
{\bfseries Computed Tomography (CT)-derived  Cardiovascular Flow Estimation Using Physics-Informed Neural Networks  Improves with Sinogram-based Training: A Simulation Study } \\  
\vspace*{10mm}
Jinyuxuan Guo, Gurnoor Singh Khurana, Alejandro Gonzalo Grande, Juan C. del Alamo, Francisco Contijoch \\
\vspace{5mm}
\vspace{5mm}
}
}

\textbf{Affiliations:} Jinyuxuan Guo (Dept. of Bioengineering, University of California San Diego), 
Gurnoor Singh Khurana (Dept. of Computer Science Engineering, University of California San Diego)
Alejandro Gonzalo Grande Dept. of Mechanical Engineering, Univ of Washington), Juan C. del Alamo (Depts of Mechanical Engineering and Cardiology, Univ. of Washington)
Francisco Contijoch (Depts. of Bioengineering, Radiology, University of California San Diego). \\

\pagenumbering{roman}
\setcounter{page}{1}
\pagestyle{plain}
\textbf{Corresponding Author Email:} fcontijoch@ucsd.edu \\

\textbf{Author Contributions:} All authors made substantial contributions to the conception or design of the work or the acquisition, analysis, or interpretation of data;
aided in drafting and reviewing the manuscript; approved the version to be published; and
agreement to be accountable for all aspects of the work.\\

\textbf{Conflict of Interest Statement:} The authors have no relevant conflicts of interest to disclose.\\

\textbf{Acknowledgments:} This study was supported in part by the US National Institutes of Health (NIH) under award number R01HL160024 and K01HL143113.\\



\newpage     
\section{Abstract}
\begin{abstract}
\noindent {\bf Background:}  Non-invasive imaging-based assessment of blood flow plays a critical role in evaluating heart function and structure. Computed Tomography (CT) is a widely-used imaging modality that can robustly evaluate cardiovascular anatomy and function, but direct methods to estimate blood flow velocity from movies of contrast evolution have not been developed.\\ 

{\bf Purpose:} This study evaluates the impact of CT imaging on Physics-Informed Neural Networks (PINN)-based flow estimation and proposes an improved framework, SinoFlow, which uses sinogram data directly to estimate blood flow.\\

{\bf Methods:} We generated pulsatile flow fields in an idealized 2D vessel bifurcation using computational fluid dynamics and simulated CT scans with varying gantry rotation speeds, tube currents, and pulse mode imaging settings. We compared the performance of PINN-based flow estimation using reconstructed images (ImageFlow) to SinoFlow.\\

{\bf Results:} SinoFlow significantly improved flow estimation performance by avoiding propagating errors introduced by filtered backprojection. SinoFlow was robust across all tested gantry rotation speeds and consistently produced lower mean squared error and velocity errors than ImageFlow. Additionally, SinoFlow was compatible with pulsed-mode imaging and maintained higher accuracy with shorter pulse widths. \\

{\bf Conclusions:} This study demonstrates the potential of SinoFlow for CT-based flow estimation, providing a more promising approach for non-invasive blood flow assessment. The findings aim to inform future applications of PINNs to CT images and provide a solution for image-based estimation, with reasonable acquisition parameters yielding accurate flow estimates. \\

\end{abstract}

\newpage    

\setlength{\baselineskip}{0.7cm}      

\pagenumbering{arabic}
\setcounter{page}{1}
\pagestyle{fancy}
\section{Introduction}
Non-invasive imaging-based assessment of blood flow plays a critical role in the evaluation of heart function and structure. 
For example, the severity of aortic stenosis is classified by estimating peak flow velocity and deriving a mean transvalvular pressure gradient
\cite{namasivayam_transvalvular_2020,otto_valvular_2006,pibarot_improving_2012}. 
In addition, advanced features of the flow pattern have been used to predict invasive hemodynamics and risk factors such as blood stasis
\cite{rodriguez-gonzalez_cardiac_2024,rodriguez-gonzalez_stasis_2025}. For example, the properties of flow vortices have been shown to correlate with mean pulmonary artery pressure
\cite{reiter_magnetic_2008,mahammedi_pulmonary_2013},
atrial thromboembolism \cite{spartera_reduced_2023}, left ventricular function \cite{martinez-legazpi_contribution_2014} and other cardiovascular conditions \cite{kheradvar_diagnostic_2019}.

Conventionally, flow maps and associated metrics are obtained using either Dopper echocardiography or phase-contrast magnetic resonance imaging (PC-MRI)
\cite{bax_2017_nodate,bermejo_clinical_2015}. 
However, X-ray computed tomography (CT) is a widely-used imaging modality that can robustly evaluate cardiovascular anatomy and function 
\cite{sun_cardiac_2012}.
Current CT systems can provide time-resolved volumetric images through a fast, volumetric, high-resolution acquisition
\cite{kelly_high-resolution_2004}.
Overall reductions in imaging dose and the capability to perform dose modulation has enables CT imaging to be performed across one or more cardiac cycles
\cite{okayama_role_2010}.
This has enabled work which evaluates the dynamics of contrast evolution through the cardiovascular system
\cite{wu_low_2025,hubbard_timing_2019}.
However, a clinically robust method to estimate blood flow velocity from movies of contrast evolution is not available yet. 
Such a method, which quantifies blood flow fields over space and time by analyzing the spatiotemporal change in intensity, could augment the anatomical and functional assessments already available and have a significant clinical impact.

Physics-Informed Neural Networks (PINNs) have demonstrated the ability to estimate fluid flow and pressure fields from spatiotemporal images of the transport dynamics of a passive tracer, such as CT contrast \cite{raissi_hidden_2020}. Estimates are obtained by training a neural network to learn an implicit representation of the observed images, while penalizing both deviations from the measured tracer concentration field and residuals of the governing physical equations, i.e., the Navier--Stokes equations, the continuity equation, and a convective transport equation.  
More recently, this inverse problem has been addressed with Optimizing a Discrete Loss (ODIL), which minimizes a loss constructed from grid-collocated discretizations of the governing equations based on standard numerical approximations (e.g., finite differences) \cite{karnakov_solving_2023}.  
These emerging frameworks can be viewed as generalizations of the classic optical flow method \cite{liu_fluid_2008}, embedding additional physical constraints beyond tracer convection and leveraging modern machine learning and optimization tools.
PINNs have also been applied for data assimilation in cardiovascular flows \cite{kissas_machine_2020,garay_physics-informed_2023}, super-resolution in 4D flow MRI \cite{fathi_super-resolution_2020,ye_multiparametric_2022}, and Doppler echocardiography \cite{maidu_super-resolution_2025,wong_3d_2025,ling_physics-guided_2024}.
However, the use of PINNs for the assessment of contrast evolutuon in CT is more limited.
While PINNs have been used to evaluate brain perfusion \cite{de_vries_accelerating_nodate}, no methods 
currently exist to reconstruct cardiovascular flow fields from 4D CT contrast. 
Notably, prior proof-of-concept studies\cite{raissi_hidden_2020} have focused on image-to-flow inference under the assumption of perfect image reconstruction without motion artifacts.

Here, in a simulation study, we evaluate the impact of CT imaging on PINN-based flow estimation and propose an improved framework that integrates the physics of flow, contrast transport, and CT image formation.
Our first hypothesis is that a PINN's ability to infer hidden flow fields from contrast dynamics is limited by by errors in reconstructed images.
CT scanners have limited spatiotemporal resolution and clinically-relevant imaging doses lead to noise in the sinogram data \cite{cao_dual-core_2017}. 
Therefore, reconstructed movies of concentration are subject to spatial blurring, motion artifacts, and noise which will be particularly pronounced when high velocities and rapid flow variations.
To test this hypothesis, we evaluate the impact of both spatiotemporal artifacts and imaging noise on estimation of flows using a PINN where images are generated using a conventional filtered back-projection (FBP) reconstruction.  
This setup, which we term ImageFlow, does not account for the fact that blood and contrast evolve during CT data acquisition. 
Therefore, a collection of views violate the the static-object assumption of FBP, leading to motion-induced artifacts in the reconstructed images. 
As these artifacts increase, we expect the accuracy of PINN-derived flow estimates to deteriorate.  
In contrast, we expect flow estimates to be more robust to imaging noise as the implicit representation of the PINN is constrained by the governing physics.
Our second hypothesis is that PINN-based flow estimation can be improved by avoiding explicit image reconstruction and instead solving directly for the concentration and flow fields from the sinogram data, a new framework we termed SinoFlow.
In SinoFlow, the forward CT acquisition process is embedded within the PINN as a flow-aware time-dependent Radon transform.
As a result, contrast motion can occur throughout CT data acquisition and will enable improved consistency between the flow physics and the measured data.
We demonstrate that SinoFlow avoids "learning" motion artifacts and enables robust estimation across various gantry rotation speeds.

Lastly, we hypothesize that sinogram-based PINN-estimation is compatible with pulsed-mode (PM) imaging, an advanced CT acquisition strategy now available on some CT scanners.
During PM imaging, the x-ray beam intensity is rapidly modulated, delivering X-rays in short "on" pulses separated by "off" periods. 
This enables lowering patient exposure without uniformly reducing overall fluence \cite{heukensfeldt_jansen_estimation_2022}\cite{wu_cardiac_2023}. 
However, the impact of PM parameters - namely, the pulse width, the number of consecutive imaging views “on” during each rotation, and duty cycle - on PINN-based flow estimation accuracy is unknown. 
We test whether SinoFlow can accurately evaluate imaging data obtained under different PMP settings to evaluate the impact of pulse width and duty cycle.  

\section{Methods}
\subsection{Generation of Flow Fields}
\begin{figure}[ht]
 \begin{center}
  \includegraphics[width=\textwidth]{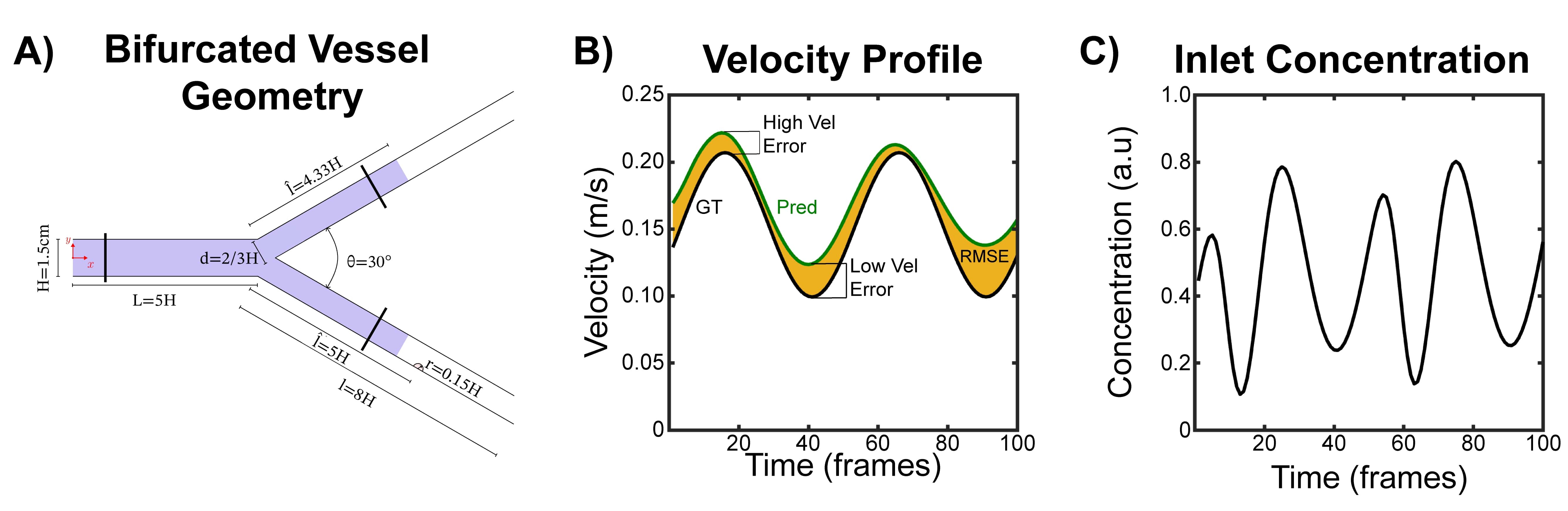}
  \caption{\textbf{Flow Simulation Geometry, Velocity Analysis, and Concentration Profile} 
  \textbf{A)} The Y-shaped tube geometry, where a clot obstructs flow in one of the outlets. Black lines mark the inlet and outlet cross-sections used for velocity and concentration measurements. 
 \textbf{B)} Velocity profile at the inlet, with the black curve representing the ground truth and the green curve showing the PINN prediction. 
 The yellow-shaded area between the two curves indicates the root mean square error (RMSE). 
 High-velocity and low-velocity errors are computed from the upper and lower 10\% of the values, respectively, while the peak-to-peak error corresponds to the difference between the 10\% high and low velocities in both the ground truth and predicted profiles. 
 \textbf{C)} Concentration profile prescribed at the inlet.}
  \label{fig_geometry}
 \end{center}
\end{figure}

We generated pulsatile flow fields throughout an idealized 2D vessel which bifurcates into two daughter vessels of the same caliber, shown in Fig.~\ref{fig_geometry}\mfig{fig_geometry}A, using computational fluid dynamics (CFD).
The parent vessel is modeled as a straight channel of height $H$ and length $L=5H$, while the daughter vessels have height $h=2H/3$ and length $l=8H$.
The angle between daughter vessels is $\alpha=30^\circ$, forming a Y-shaped bifurcation symmetric with respect to the parent vessel centerline.
Importantly, flow split between the daughter vessels is imbalanced as one daughter vessel is simulated to be obstructed by a downstream thrombus or stenotic region. 
This feature will be reflected in the concentration evolution movie but is unknown from the anatomy in the region of interest (blue region in Fig \ref{fig_geometry}A).
The flow within the vessel at the inlet is shown in  Fig.~\ref{fig_geometry}\mfig{fig_geometry}B.
The occlusion in the lower daughter vessel is modeled with a semicircle of radius $r = 0.15H$, resulting in a maximum occlusion of $45\%$ of the diameter ($d$). 
Addition details regarding the CFD boundary conditions are described in Appendix \ref{subsec:appendix_cfd_bc}

\subsection{Prescription of Contrast Agent Concentration}
Our starting point to model contrast agent concentration dynamics was simulated the convection-diffusion equation for a passive scalar carried by the flow,
\begin{equation*}
\frac{\partial c_{\text{\tiny CFD}}}{\partial \tilde{t}} + \nabla \cdot \left( \mathbf{\tilde{u}} c_{\text{\tiny CFD}}\right) = \frac{1}{Pe} \nabla^2 c_{\text{\tiny CFD}},
\label{eq_passive_scalar}
\end{equation*}
where $Pe = u_c H / D_{c_{\text{\tiny CFD}}}$ is the P\'{e}clet number that measures the relative importance of convection over diffusion, and $D_{c_{\text{\tiny CFD}}}$ is the mass diffusion coefficient.  
Using this value together with the characteristic velocity $u_c$ and vessel height $H$ of the idealized bifurcation gives $Pe \sim 10^{7}$, indicating that diffusion is negligible. 
Therefore, the evolution of Computational fluid dynamics (CFD)-derived concentration fields, $c_{\text{\tiny CFD}}$ is governed by pure advection, so we model it as
\begin{equation*}
\frac{\partial c_{\text{\tiny CFD}}}{\partial \tilde{t}} + \nabla \cdot \left( \mathbf{\tilde{u}} c_{\text{\tiny CFD}}\right) = 0.
\label{eq_cca}
\end{equation*}

To solve this first-order partial differential equation, it is sufficient to prescribe inlet 
boundary conditions at the parent vessel, $c_{\text{\tiny CFD}}(x=0,y,t)$. 
In our model, we impose a spatially uniform profile, $c_{\text{\tiny CFD}}(x=0,y,t) = C_0(t)$, as illustrated in 
Fig.~\ref{fig_geometry}\mfig{fig_geometry}C.  
Additional details regarding the numerical scheme used to solve for the concentration are provided in Appendix \ref{subsec:appendix_cfd_concentration}

Of note, the solution domain for $c_{\text{\tiny CFD}}$ (blue region in Fig.~\ref{fig_geometry}A) was deliberately restricted to exclude the daughter vessel narrowing, in order to mimic a scenario where the anatomical cause of the flow imbalance is absent from the CT images.

\subsection{CT Scan Simulation}
\label{sec:scan_simulations}
CFD-derived concentration fields, $c_{\text{\tiny CFD}}(x,y,t)$, were used as the basis to simulate CT scanning. 
The concentration $c_{\text{\tiny CFD}}$ was normalized to range from 0 to 1, with $t$ denoting 100 temporal frames spanning two cardiac cycles. 
From these fields, we generated $1600\, px \times  1600\, px$ images covering a $50\, cm \times 50\,cm$ region to mimic the imaging field-of-view of currnet CT scanners.
The vessel was located centrally and had a diameter of $1.5\,cm$, approximating the size of the human inferior vena cava (IVC).

Fan-beam imaging was simulated in Python with user-defined gantry rotation speed (GRS) at 1, 2, 3, 
4, 6, 8, and 10 Hz. 
This range reflects the capabilities of current clinical CT scanners, allowing us to compare the ImageFlow and SinoFlow PINN frameworks in images with motion artifacts of varying severity. 
We simulated 1600 detector elements, each integrating 5 x-ray beams. 
The geometry was set to use a fan angle of 43.6 degrees and 984 views were acquired per gantry rotation.
We simulated different gantry starting angles (0, 30, 60, 90, 120 and 150 degrees) and averaged the results since different gantry positions lead to different motion artifacts and the starting angle is unknown and arbitrary at the time of scanning \cite{contijoch_impact_2017}.

\subsubsection{Addition of Noise to Sinogram Data}
\label{sec:Pois_nois}
Imaging noise was modeled by adding Poisson-distributed fluctuations to the obtained sinogram values.
The x-ray attenuation through a non-homogeneous material was computed using the Beer–Lambert law,
\begin{equation}
I = I_0 \, \exp\left(-\int_0^L \mu(x) \, dx\right)
\end{equation}
where $I$ is the intensity of the X-ray beam after passing through the material, 
$I_0$ is the incident X-ray intensity, $\mu(x)$ is the linear attenuation coefficient at position 
x along the beam path, $x$ represents the position variable, and $L$ is the total length the X-
ray travels through the object. 
Noisy measurements were then obtained as
\begin{equation}
I_n \sim {\rm Poisson}(I)
\end{equation}
The initial X-ray intensity, $I_0$, was varied from $10^5$ to $10^{15}$ to generate images with varying contrast-to-noise ratios, thereby mimicking a range of CT dose levels from low- to high-dose imaging.

\subsubsection{Filtered Backprojection Reconstruction}
Images were reconstructed using a standard filtered backprojection (FBP) algorithm with a RAM-LAK filter \cite{demirkaya_reduction_nodate}.
This reconstruction yielded {\it imaged} concentration fields $\ c_{\text{\tiny CT}}(x, y, t)$, represented as 
time-resolved movies, which approximate but do not exactly equal the true concentration. 
The movies had different amounts of motion blurring due to the range of gantry rotation speeds and initial x-ray intensity value, $I_0$, which controlled the contrast-to-noise ratio.

\subsection{Physics-informed Neural Network Architecture}
\begin{figure*}[ht]
    \centering
    \includegraphics[width=\linewidth, page=1]{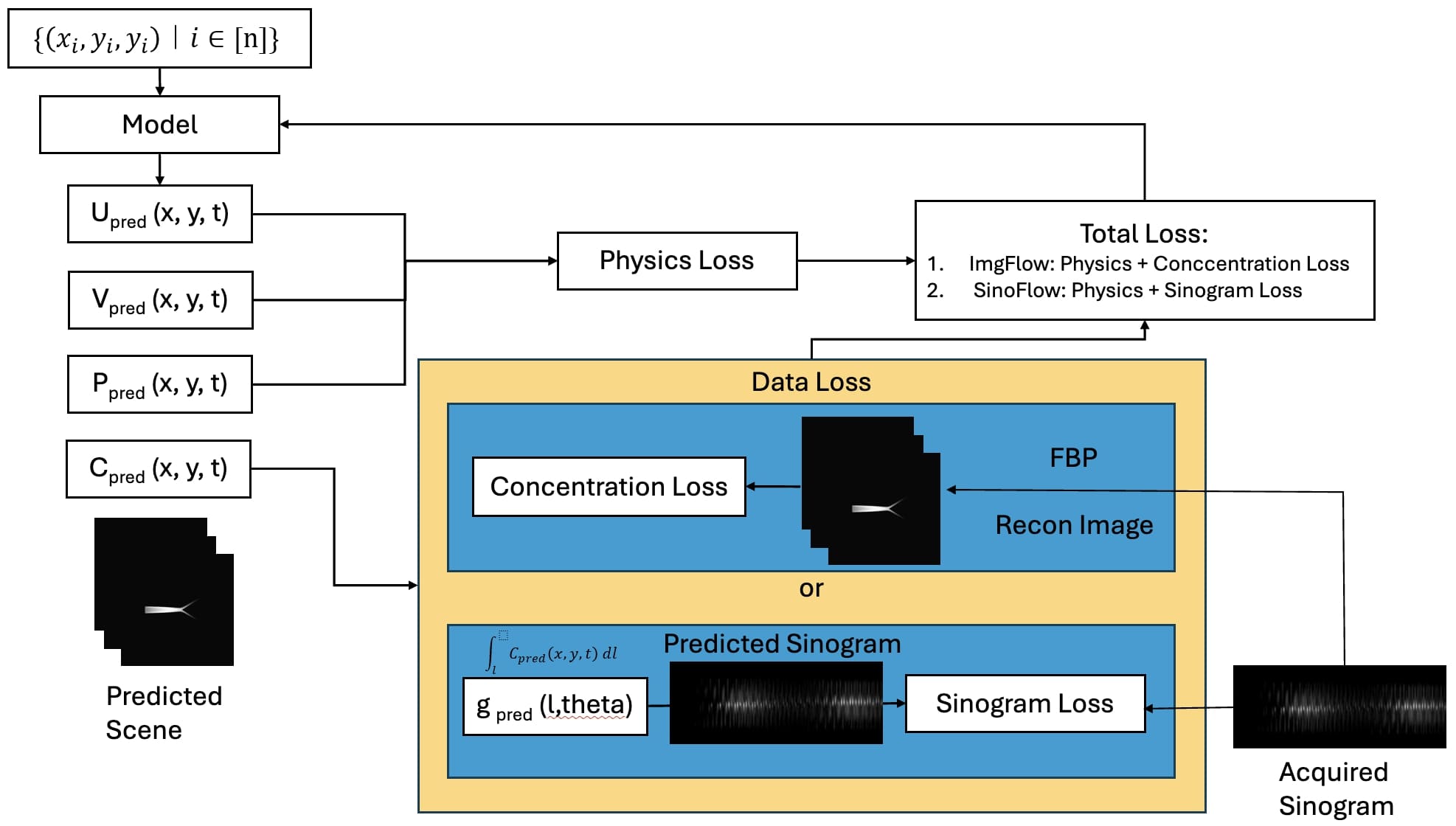}
    \caption{
    \textbf{PINN-based Flow Estimation Approaches}
    In both ImageFlow and SinoFlow, the same image domain PINN is used to predict concentration and flow field movies. 
    However, SinoFlow includes the forward rendering process in the pipeline to generate a predicted sinogram $g_{pred}$ and therefore calculates a sinogram loss as the data loss, while ImageFlow compares generated concentrations $c_{pred}$ to FBP-reconstructed concentration values.}
    \label{fig_pipeline}
\end{figure*}

In all cases, a Multi Layer Perceptron (MLP) network with 10 hidden layers and 200 neurons per layer and Sigmoid Linear Unit (SiLU) activation function were used to create an implicit representation of the concentration field in the contrast evolution as a function of time time\cite{elfwing_sigmoid-weighted_2018}. 
This model was implemented in Python using the PyTorch library. 
Neuron coefficients were determined using a Adam optimizer over $3\times10^5$ iterations with a learning rate of $10^{-3}$. 
This framework, generalized for ImageFlow and SinoFlow, is visualized in Figure \ref{fig_pipeline}\mfig{fig_pipeline}, with further details provided below.
These and other hyperparameters, such as the loss weight coefficients and the number of sampling points, were selected heuristically and tuned experimentally to balance the contributions of sinogram data and flow information in model supervision.

\subsubsection{ImageFlow}
\label{sec:imageflow}
The input to ImageFlow is $(t, x, y)$, where $t$ represents the non-dimensional time elapsed in 
the contrast evolution movie, while $x$ and $y$ are the non-dimensional spatial coordinates.
The network outputs a quadruplet of values $(\hat c, \hat u, \hat v, \hat p)$ representing the concentration,  $x-$ and $y-$velocity components, and pressure at the specified spatiotemporal positions.
The PINN weights are computed by minimizing a loss function that aggregates data loss and 
residuals from physical constraints, 
\begin{equation}
\mathcal{L} = \mathcal{L}_{\text{physics}} + \mathcal{L}_{\text{data}},
\end{equation}

In ImageFlow, the physical loss is the L2 norm of the residuals of partial differential equations 
governing contrast transport, mass conservation, and fluid momentum balance (i.e., the Navier-Stokes equations)\cite{raissi_hidden_2020} whereas the data loss is the L2-norm between inferred and {\it imaged} contrast agent concentration,
\begin{equation}
    \mathcal{L}_{data} = \lambda_0 \mathcal{L}_{\text{conc}} = \lambda_0 \frac1n \sum_{n} \left[\hat c(x,y,t) - \ c_{\text{\tiny CT}}(x,y,t)\right]^2,
\end{equation}
and $\lambda_0$ is a hyperparameter controlling the relative weight of data misfit and 
physical constraints.
More details are provided in Appendix \ref{subsec:appendix_pinn_losses})
Both loss function terms are evaluated at a set of points randomly sampled from the spatiotemporal
domain of CFD simulation.
In ImageFlow, $\lambda_0$ was set to 1. 
%

\subsubsection{SinoFlow}
\label{sec:sinoflow}
%
In SinoFlow, we modified the PINN architecture to learn the $(x,y,t) \longrightarrow (\hat c, \hat u, \hat v, \hat p)$ representation from the the CT sinogram, i.e., the original measurement domain.
This avoids the propagation of motion artifacts generated by filtered back-projection.
We hypothesized that this would mitigate imaging artifacts (in this case, motion artifacts) introduced during the image reconstruction processes. 
To do so, the forward rendering operation, i.e., the Radon transform, was added to the inference 
framework as a hard constraint based on the time-evolving PINN-inferred concentration field,
\begin{equation}
\hat g(s, \theta) = \int_{L(\theta)} \hat c[x(s,\theta), y(s,\theta), t] \, dl
\end{equation}
where $\hat g(s,\theta)$  and $\hat c$  is the inferred sinogram and inferred concentration field. $[x(s,\theta), y(s,\theta)]$ represents the projection ray for a certain position $s$ and gantry angle $\theta = \theta_0 + \omega t$, with $\omega$ being the gantry rotation speed. 
To be compatible with forward rendering, a set of collinear points corresponding to randomly sampled locations on the sinogram were sampled during training (instead of random spatiotemporal locations as done by ImageFlow).

This architecture allows for the data loss to be redefined as the difference between the inferred sinogram $\hat g(s,\theta)$ and the measured sinogram $g(s,\theta)$, 
\begin{equation}
\mathcal{L}_{\text{data}} 
= \lambda_1 \mathcal{L}_{\text{sino}} 
= \frac{\lambda_1}{n} \sum_{n} \left[\hat g(s, \theta) - {g}(s, \theta)\right]^2.
\end{equation}

To balance data and physics loss magnitudes,$\lambda_1$ was set to $1 / n_p$, where $n_p$ is the number of spatial points used to calculate $g(s,\theta)$. In our experiments $n_p$ was equal to 1600. 
%
\subsection{Description of Computational Experiments}

\subsubsection{Impact of Gantry Rotation Speed on Image and Sinogram-based Flow Estimation}
\label{sec:exp_gantryrot}

We compared the performance of ImageFlow to SinoFlow (described respectively in \S~\ref{sec:imageflow} and \S~\ref{sec:sinoflow}) 
for scan simulations generated using clinical scanner geometry for varying rotation speeds(\S~\ref{sec:scan_simulations}). 
The impact of gantry rotation speed and starting position on resulting FBP image quality was measured by mean-square-error.
Then, the quality of the concentration fields $\hat c(x,y,t)$ estimated by ImageFlow and SinoFlow were compared to the ground truth, 
$c_{\text{\tiny CFD}}(x,y,t)$. 
ImageFlow- and SinoFlow-based flow estimaton was evaluated by measuring the accuracy of inferred velocity components
$[\hat u(x,y,t),\hat v(x,y,t)]$ at the inlet and outlets via comparison to ground truth CFD-generated data $[u(x,y,t), v(x,y,t)]$.

\subsubsection{Robustness of ImageFlow and SinoFlow to CT Imaging Noise}
\label{sec:noise_exp}
%
%
The experiments in \S~\ref{sec:exp_gantryrot} focused on the effect of gantry rotation speed (GRS) on motion artifacts, assuming noise-free data. 
In that setup, increasing GRS reduces motion blur but also requires a higher imaging dose to maintain the same photon statistics per view. 
Conversely, if $I_0$ is held constant across rotation speeds, dose remains unchanged but projection noise increases with GRS because each projection is acquired over a shorter exposure time. 
To account for these dose–noise trade-offs, we performed experiments where we modeled the imaging noise by adding Poisson-distributed fluctuations to the sinogram data (see \S~\ref{sec:Pois_nois}). 
The resulting reduction in x-ray fluence (lower CNR) was assessed in terms of its effect on both the reconstructed contrast concentration and velocity fields, as in the previous section. 
Likewise, the performance of both ImageFlow and SinoFlow was compared.
To isolate the effect of noise independently from motion blur, all scans were performed at a fixed gantry rotation speed of 4 Hz.

\subsubsection{Pulsed Mode Imaging}
\label{sec:pulse_mode}
Finally, we evaluated the compatibility of SinoFlow with pulse mode (PM) imaging, a modality in which
x-ray acquisition occurs in discrete intervals interleaved with periods without data. 
To simulate this acquisition mode, the PINN data loss $\mathcal{L}_{data}$ was limited to portions of the sinogram during which the pulses were obtained.  
Experiments were performed varying the PM duty-cycle (10 to 100 percent) for 10- and 50-view pulse widths, where the pulse width is the number of consecutive views that are obtained between non-imaging intervals. 
The 10-view pulse was the shortest pulse duration tested.
This was selected such that even a high duty cycle (such as 75\% which corresponds to $\sim$3 "off views") is feasible given previously reported pulse rates \cite{haneda_bolus_2025}.

\subsubsection{Analysis of Flow Fields and CT Parameters}
\label{sec:strouhal_analysis}
To support the definition of flow fields and CT gantry rotation speeds we prescribed, we can consider a simplified one-dimensional advection model. 
Over one gantry rotation, $\Delta t = 2\pi/\omega_{\text{gantry}}$, a first-order Taylor expansion 
with the advection relation ($\partial_t c = -u\,\partial_x c$) gives the mismatch
\[
\Delta c \;\sim\; \frac{2\pi u}{\omega_{\text{gantry}} L_c},
\]
where $L_c$ denotes the bolus-front thickness along the vessel 
centerline. 
Motion artifacts are negligible when this term is small, which yields
\(
\omega_{\text{gantry}} \gg {2\pi u}/{L_c}.
\)
Using the flow Strouhal number
\(
St_{\text{flow}} = {\omega_{\text{flow}} H}/{u},
\)
we obtain
\[
\frac{\omega_{\text{gantry}}}{\omega_{\text{flow}}} \gg 
\frac{2\pi}{St_{\text{flow}}}\,\frac{H}{L_c}.
\]
In our Y-shaped anatomy, $St_{\text{flow}} \approx 0.37$, $\omega_{\text{flow}} \approx 7.3 
s^{-1}$, and $L_c/H \approx 5$, implying a threshold of $f_{\text{gantry}}\sim 3.8$~Hz.
This is close to the 4 Hz imaging scenario (which is a common GRS with current scanners). 
More broadly, the Strouhal-based framework may offer a general basis for predicting threshold behavior across different flow conditions and geometries.

\subsection{Statistical Analysis}
\label{sec:stats_analysis}
Accuracy of the reconstructed concentration images and the predicted concentration fields from ImageFlow and SinoFlow was assessed using root mean-square error (RMSE) against the ground truth CFD-generated concentration fields. 
Flow accuracy at the inlet and outlets was quantified using RMSE together with peak velocity, minimum velocity, and velocity range (peak–minimum) at the inlet and outlet locations. 
In addition, the ability to detect the small outlet flow imbalance was evaluated using the ratio of outlet flows (Outlet Ratio) and compared to the CFD-generated ratio. 

A two-way repeated-measures analysis of variance (RM-ANOVA) was conducted to assess the effects of GRS, CNR, or Duty Cycle and neural network framework (ImageFlow and SinoFlow) or PM views on metrics of interest. 
Given each condition had n=6 observations (corresponding to different gantry starting angles), we did not test for normality.
However, ANOVA's have been shown to be robust against small variations in normality \cite{blanca_non-normal_2017}.
For significant main effect or interaction, post-hoc pairwise comparisons were conducted using paired t-tests, with Bonferroni correction applied to adjust for multiple comparisons.
Additionally, two-sample t-tests were performed to compare reconstruction results against the ground truth (when available) and between the two neural network frameworks. 
Statistically significant differences ($p < 0.05$) are indicated by an asterisk (*) in the corresponding figures.

\section{Results}

\subsection{Impact of GRS on flow estimation}
\begin{figure*}[ht]
    \centering
    \includegraphics[width=\linewidth, page=1]{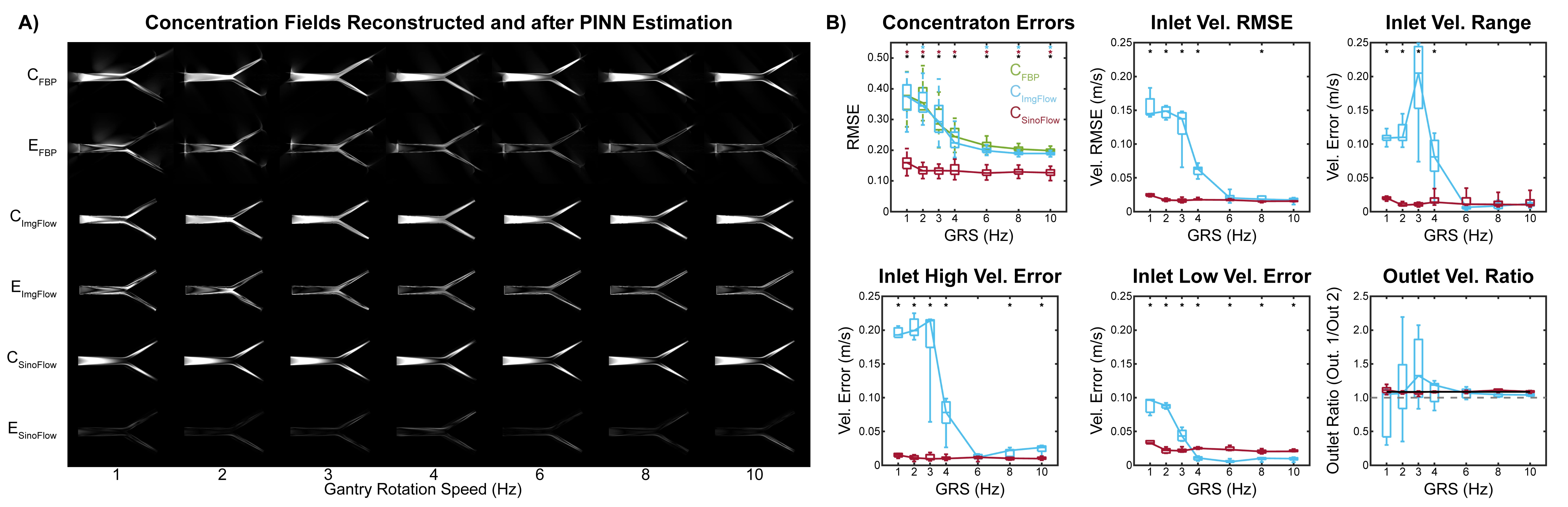}
    \caption{
    \textbf{FBP and PINN-predicted concentration scenes and boxplots of ImageFlow and SinoFlow results at different GRS}.
    \textbf{A)} The FBP, ImageFlow predicted and SinoFlow predicted concentration scenes and their differences compared with the ground truth at 1, 2, 3, 4, 6, 8, and 10 Hz GRS. 
    \textbf{B)} ImageFlow (blue) and SinoFlow (red) predicted concentration and flow velocity results at the inlet, using boxplots (each box contains 6 data points from the 6 tested gantry starting angle), including concentration RMSE, velocity RMSE, high velocity error, low velocity error, velocity range error and outlet velocity ratio across GRS. 
    FBP concentration errors are shown in green. 
    The grey dashed line in the outlet velocity ratio boxplot indicates the ratio of 1 (equal flow between the two outlets), and the black line the ground truth ratio. 
    For statistical analysis, black asterisks indicates significant difference found between ImageFlow and SinoFlow results while blue asterisks compares FBP and ImageFlow and red asterisks compares FBP and SinoFlow.}
    \label{fig_GRS}
\end{figure*}

Figure~\ref{fig_GRS}\mfig{fig_GRS}A shows the FBP-reconstructed concentration field, $c_{\text{\tiny FBP}}$, together with the PINN-predicted concentrations from ImageFlow ($\hat c_{\rm ImageFlow}$) and SinoFlow ($\hat c_{\rm SinoFlow}$) at different GRS. 
For each case, the absolute concentration error, relative to the ground-truth concentration from CFD simulations $c_{\text{\tiny CFD}}$, is visualized. 
The images demonstrate that motion artifacts in reconstructed CT images increase as GRS decreases, and that the spatial distribution of errors also depends on GRS. 
At the lowest GRS values, where overall errors are larger, the discrepancies appear concentrated near the vessel walls and in the bifurcation region. 
ImageFlow and FBP yield similar concentration fields, whereas SinoFlow consistently produces lower 
errors across all GRS.
These findings were confirmed in quantitative assessment, shown in Figure \ref{fig_GRS}\mfig{fig_GRS}B. 
Concentration RMSE increased from $\approx 0.2$ at 8-10 Hz to $\approx 0.375$ at 1Hz for both FBP and ImageFlow. 
At 1,2, and 4 Hz, there was no statistical difference in RMSE between FBP and ImageFlow. This suggests the errors in FBP lead to errors in concentration fields predicted by ImageFlow (shown in blue).
This similarity is to be expected, since the ImageFlow PINN was trained on the FBP-reconstructed concentration fields.

Errors in FBP reconstructed images also appear to propagate into errors in velocity estimates derived from ImageFlow.
Inlet results are shown in Figure \ref{fig_GRS}\mfig{fig_GRS}B while Supplemental Figure \ref{suppfig_GRS_Outlet}\mfig{suppfig_GRS_Outlet} demonstrates errors in outlet estimates.
Inlet velocity RMSE, velocity range, high velocity error, and low velocity error increased sharply as GRS decreased beyond 4 Hz.
The outlet ratio (used for detection of the small flow imbalance between the daughter vessels) could be estimate when imaging errors were small. 
The CFD-derived Outlet Ratio is 1.086.  
At GRS $\ge$ 4 Hz, ImageFlow predicts a small imbalance (1.036) which is close to the correct value.
However, Figure \ref{fig_GRS}\mfig{fig_GRS}B shows how estimates become highly variable and inaccurate for GRS $\leq$ 4 Hz.

SinoFlow predictions were more accurate and robust with respect to gantry rotation speed (Fig. \ref{fig_GRS} A and B).
The concentration fields inferred by SinoFlow had RMSE of $\approx 0.133$.
This was significantly lower than the RMSEs obtained with FBP or ImageFlow, even when comparing SinoFlow at $GRS=1$ Hz to ImageFlow at the highest rotation speed, $GRS=10$ Hz.
Further, the difference in RMSE between $GRS=10$ Hz (0.189) an $GRS=1$ Hz (0.375) was small.
SinoFlow also produced stable and accurate flow curves. 
Specifically, velocity RMSE, velocity range error, high velocity error, and low velocity error all remained near 0.01 - 0.03 m/s regardless of GRS. 
Moreover, Outlet Ratio predicted by SinoFlow (1.092) more closely matched the ground-truth value (1.086) and exhibited markedly less variability than those obtained with ImageFlow.

Table \ref{tab:grs_results}\mtab{tab:grs_results} provides numerical values of the evaluation metrics for GRS = 4 Hz.

\subsection{Impact of imaging noise on flow estimation}
\begin{figure*}[ht]
    \centering
    \includegraphics[width=\linewidth, page=1]{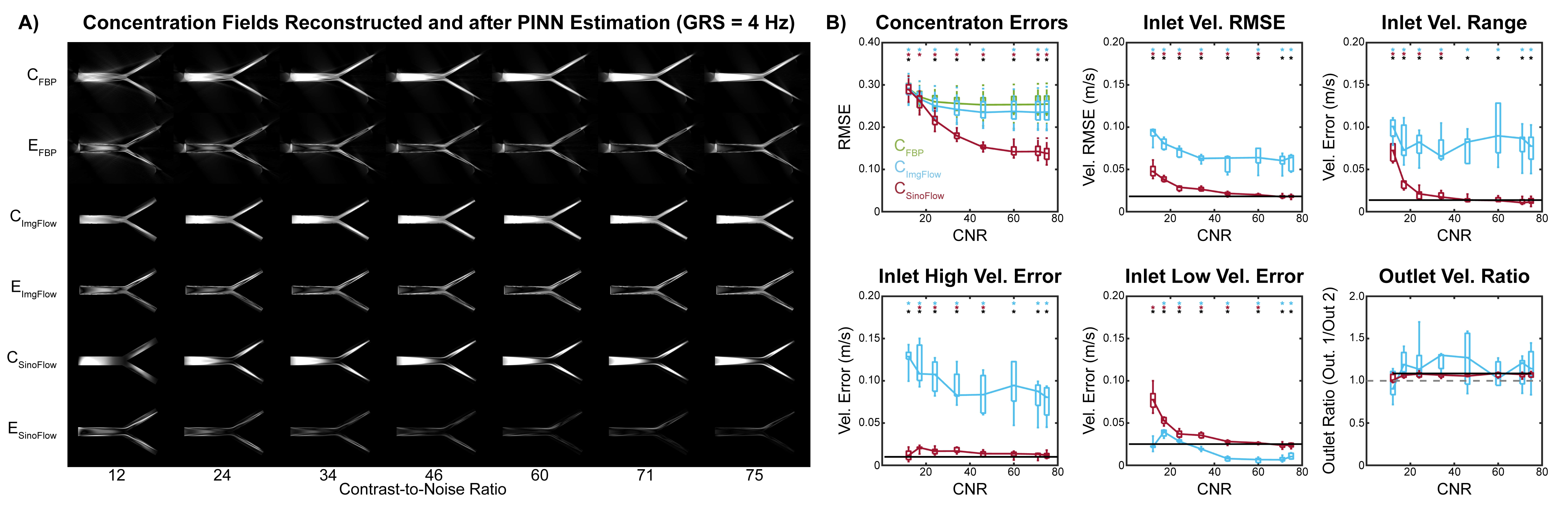}
    \caption{
    \textbf{FBP and PINN-predicted concentration scenes and boxplots of ImageFlow and SinoFlow results at different CNR.}
    \textbf{A)} FBP, ImageFlow predicted and SinoFlow predicted concentration scenes and their differences compared with the ground truth at 4 Hz GRS and 12, 24, 34, 46, 60 71 and 75 CNR. 
    \textbf{B)} Concentration and flow velocity results at the inlet  at 4 Hz GRS and varying CNR. FBP results are shown in green while ImageFlow are in blue and SinoFlow results are in red. 
    Black lines for the four velocity error panels are the mean noise-free SinoFlow results at 4 Hz. 
    The black line for the outlet ratio is the ground truth ratio.    
    For statistical analysis, black asterisks indicates significant difference found between ImageFlow and SinoFlow results while blue asterisks compares FBP and ImageFlow and red asterisks compares FBP and SinoFlow.}
    \label{fig_noise}
\end{figure*}

Figure \ref{fig_noise}\mfig{fig_noise}A shows how increasing sinogram noise leads to artifacts in the concentration fields reconstructed by FBP, ImageFlow, and SinoFlow.
At high CNR, artifacts were largely confined to near-wall regions and, in the case of FBP and ImageFlow, also appeared in the inlet region. 
As CNR decreased, these near-wall and inlet errors intensified, and additional artifacts emerged within the parent vessel lumen.
The relationship between image CNR and concentation field RMSE is shown in Figure \ref{fig_noise}\mfig{fig_noise}B (top left panel).
At low CNR (CNR=12), all three methods yielded similar errors (RMSE$\approx$0.29). 
With increasing CNR, SinoFlow exhibited a more pronounced improvement, achieving an RMSE of 
0.142 at CNR=75, significantly lower than imageFlow (0.237) and FBP reconstructed
images (0.253). 

SinoFlow also resulted in better velocity error metrics, both at the inlet (Figure \ref{fig_noise}\mfig{fig_noise}B) and outlet (Supplemental Figure \ref{suppfig_Noise_Outlet}\mfig{suppfig_Noise_Outlet}).
At high CNR (CNR=75), the inference errors observed after the addition of sinogram 
noise were similar to those obtained under noise-free imaging. 
As CNR decreased, velocity RMSE, velocity range error, and low velocity error progressively 
increased. At low CNR (CNR=12), these metrics reached 0.045, 0.073, and 0.075 m/s respectively,  
corresponding with $\sim5x$, $\sim5x$, and $\sim3x$ higher values than under noise-free imaging.
Notably, the high velocity and outlet ratio remained accurate and showed almost no dependence on CNR.

Overall, Imageflow underperformed relative to SinoFlow.
Velocity RMSE, velocity range, high velocity error, and outlet ratio error were all higher for 
ImageFlow than SinoFlow, and outlet ratio predictions were generally less accurate.   
However, low-velocity error was consistently better for ImageFlow across all tested CNR levels.

\subsection{Accuracy of SinoFlow during PM imaging}
\begin{figure*}[ht]
    \centering
    \includegraphics[width=\linewidth, page=1]{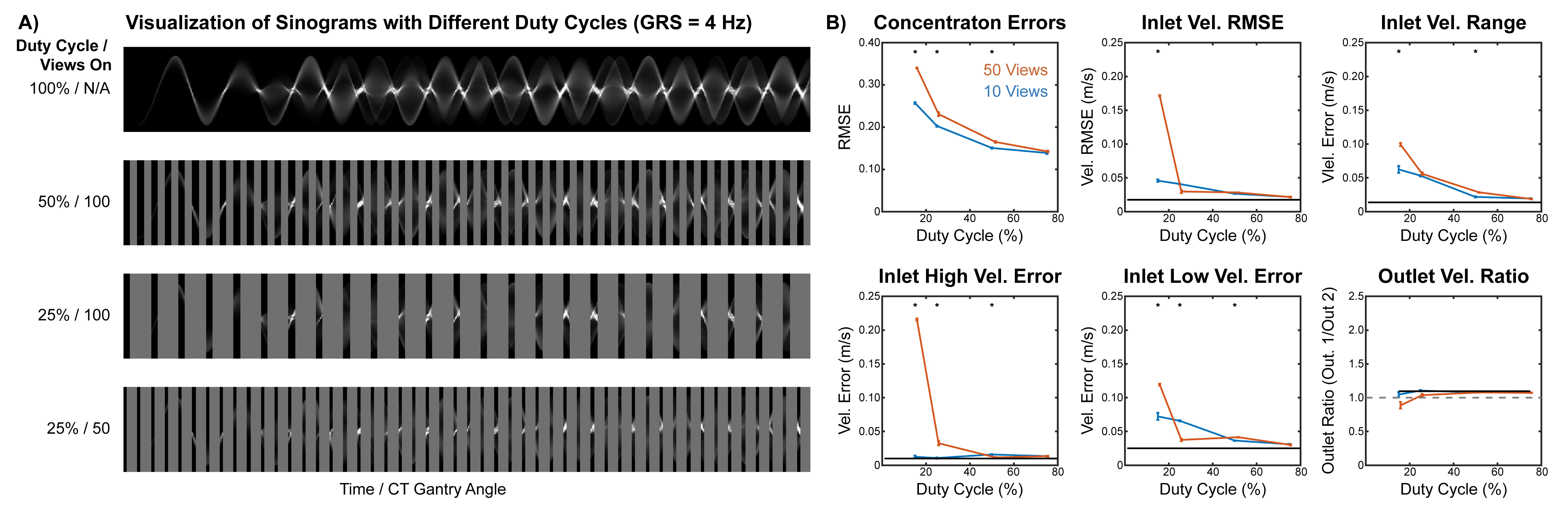}
    \caption{
    \textbf{Pulse Mode Imaging Sinograms and Performance of SinoFlow at different duty cycles and pulse widths.}
    \textbf{A)} Visualization of pulse mode imaging sinogram at 100\%, 75\% and 50\% duty cycle (grey areas on the sinogram were skipped in PINN solving) at 4 Hz GRS and CNR=60. The last row uses a smaller pulse width, causing a shorter on and off period. 
    \textbf{B)} SinoFlow predicted concentration and flow velocity results at the inlet at 15\%, 25\%, 50\% and 75\% duty cycle. 10-view (blue) and 50-view (orange) pulse widths were evaluated. 
    Black lines for the four velocity error panels are the mean SinoFlow results at 4 Hz GRS and 60 CNR. 
    The black line for the outlet ratio is the ground truth ratio.    
    For statistical analysis, black asterisks indicates significant difference found between between 10-view and 50-view pulse width results.}
    \label{fig_duty}
\end{figure*}

The accuracy of SinoFlow under pulsed mode imaging was studied by characterizing the PINN's 
inference accuracy as a function of duty cycle and pulse width.
As in previous sections, error metrics were assessed at the inlet (Figure \ref{fig_duty}\mfig{fig_duty}) and outlet of the flow model (Supplemental Figure \ref{suppfig_DutyCycle_Outlet}\mfig{suppfig_DutyCycle_Outlet}).
As duty cycle decreased, all four error metrics increased. 
However, inferences using a 10-view pulse width consistently outperformed those using a 50-view pulse width, with particularly pronounced differences at the lowest duty cycle tested. 
This result highlights the importance of the number of pulses: for the same duty cycle, a 10-view pulse width produces more pulses than a 50-view pulse width, leading to denser temporal sampling and improved PINN estimation accuracy.

At a 75\% duty cycle, SinoFlow showed no statistically significant differences between 10-view and 50-view pulse widths across any of the four velocity error metrics. 
However, at 15\% duty cycle, the effect of pulse width became evident.
The high velocity error from 10-view pulses was 0.015 m/s which is comparable to the 100\% duty cycle scenario whereas for 50-view pulses, it reached 0.19 m/s, ($>$10x higher). 
The low velocity error, peak-to-peak error, and RMSE followed similar trends, reaching 0.12, 0.097 and 0.17 m/s for 50-view pulse width at 15\% duty cycle. 
This is roughly 70\%, 58\%, and 267\% higher than the corresponding from 10-view pulse errors.

For duty cycles above 50\%, the predicted outlet ratio was accurate for both 10-frame and 50-
frame pulse widths, with values around 1.079. 
At a 25\% duty cycle, the outlet ratio from 10-view pulses remained close to the ground truth at 1.095, whereas 50-view pulses produced a less accurate value of 1.038. 
When the duty cycle dropped further to 15\%, accuracy declined for both cases, yielding outlet ratios of approximately 1.046 (10-view) and 0.887 (50-view).

\section{Discussion}
This study demonstrates that CT acquisition parameters, particularly gantry rotation speed and imaging tube current, strongly influence the accuracy of PINN-based flow estimation from CT contrast dynamics. 
We compared two different PINN frameworks - ImageFlow, which learns from FBP-reconstructed contrast movies, and SinoFlow, which integrates sinogram data directly - and demonstrate how the sinogram-based approach leads to improve flow estimation presumably due to the fact that it limits the effects of motion artifacts present in FBP images. 

Prior work has shown that PINNs can infer flow from the dynamics of a passive tracer under idealized assumptions \cite{raissi_hidden_2020,garay_physics-informed_2023}. 
However, these studies sidestepped the impact of imaging artifacts and noise that inevitably arise in practice. 
Addressing this limitation is essential for clinical translation. 
Our results provide the first systematic evidence that PINN inference from CT contrast dynamics benefits substantially from moving beyond image-domain representations. 
By enforcing consistency between acquisition physics and tracer transport directly in the sinogram domain, SinoFlow achieves robust performance even under challenging conditions, such as slower gantry rotation speeds or reduced-dose protocols, whereas ImageFlow-based flow estimation accuracy deteriorates.  

We evaluated SinoFlow and ImageFlow using a CFD model of flow through a bifurcating vessel with partial downstream obstruction, used here as an exemplar case.
Gantry rotation speeds below $\approx 4$ Hz produced pronounced motion artifacts in reconstructed CT images, leading to higher mean squared error and velocity inaccuracies that propagated into ImageFlow estimates. 
This observation is consistent with the threshold predicted under the assumption that contrast remains nearly steady during advection over a single gantry revolution (Section \ref{sec:strouhal_analysis}). This finding suggests that the influence of motion artifacts on image-based flow estimation can be reasonably approximated from the relative gantry and flow timescales, although such estimation may become more challenging in complex geometries (e.g., aneurysms or cardiac chambers).

In contrast, SinoFlow remained robust beyond the predicted 4 Hz cutoff and consistently yielding lower concentration errors and more accurate velocity fields.
These results suggest that SinoFlow may better support CT-based flow estimation in practice, since it maintained accuracy in our benchmark case where vessel bifurcation and obstruction produced rapid contrast dynamics that amplified motion artifacts in conventional reconstructions.

While gantry rotation speed was a major determinant of motion artifacts and flow estimation 
accuracy, other imaging parameters also affect PINN performance.
To evaluate the role of imaging dose, we conducted two complementary experiments: varying imaging noise and 
simulating pulsed-mode (PM) imaging. 
This aimed to evaluate practical considerations for optimizing CT-based flow measurements.
We found that increasing imaging noise degraded flow estimation accuracy when using ImageFlow, despite the denoising capacity of the PINN framework. 
Decreasing CNR also led to increasing errors  in the reconstructed concentration field for SinoFlow.
However, SinoFlow provided higher flow estimate accuracy.
This suggests that an additional benefit of performing sinogram-based training, over image-based assessment.

We also found that SinoFlow is compatible with pulse mode imaging but that choice of parameters impacts performance.
Specifically, shorter pulse widths (10 views) maintained higher accuracy than longer pulse widths (50 views), particularly at lower duty cycles. 
This suggests that higher temporal sampling density achieved with shorter pulses provides better information than longer periods of "continuous sampling". 
Given that both scans yield the same imaging dose, it appears the benefit stems from frequent information about contrast dynamics (or lack of "blindspots" during imaging).

These findings aim to inform future application of PINNs to CT images and provide a solution (SinoFlow) which avoids errors observed with image-based estimation. 
For specific applications, the extent of motion artifact (ratio of flow to gantry rotations speed) and the impact of imaging noise should be carefully evaluated.

Our study had several limitations. 
First, we evaluated a single, idealized bifurcating geometry aligned with the axial imaging plane. 
Flow estimation in vessels oriented orthogonal to this plane (e.g., along the $z$-direction) may involve different imaging considerations, and errors may also vary in more complex vascular or cardiac chambers. 
The Y-shaped model was chosen deliberately for its simplicity, while still embedding hidden information in the imaging data.
Specifically, the flow imbalance was introduced by an obstruction located outside the field of view.
Thus, it would not be apparent to CFD-based approaches or analysis based on the geometry alone.
Therefore, this design provided a tractable platform for analysis while ensuring that accurate inference required sensitivity to contrast dynamics rather than purely geometric cues. 
Future work will extend the framework to evaluate flows along the $z$-direction and in more physiologically realistic vascular anatomies.

Second, we performed 2D imaging using a fan-beam geometry. 
This simplification enabled the evaluation of key CT parameters while avoiding significant increases in computational complexity. 
Specifically, while the PINN architecture is able to solve for 3D concentrations and flows \cite{raissi_hidden_2020}, doing so would necessitate higher computational demands, making it difficult to perform the large number of independent inferences presented here to evaluate PINN performance vs. parameters like GRS, sinogram noise, pulse mode duty ratio, and pulse width.
Given the promising results here, extension to 3D imaging is planned as future work.

Third, we intentionally designed the contrast injection to have high spatiotemporal variations to enable successful PINN-based flow estimation. 
Because PINNs infer velocity from the evolving tracer field, frames or locations in which contrast is absent or uniformly distributed yield little to no usable flow information. 
To mitigate this, we imposed an oscillatory injection profile that generated graded concentration gradients across the vessel. 
In practice, however, achieving such patterning may be challenging.
The degree of oscillation that can be introduced is limited by clinical constraints, and as the bolus travels from the injection site to the imaging region it will undergo mixing and dispersion. 
A systematic investigation of the sensitivity of PINN-based flow estimation to injection protocols will be an 
important direction for future work.
The impact of different injection protocols and the ability to be realized in practice is left for future work.

One possible improvement in this area could come from dual-energy or photon-count CT systems.
In both methods, iodine images/data can be generated. 
This data would be better for PINN-estimation than the conventional sinogram or images which mix contrast agent concentration data with "background" intensity values.
By supplying higher fidelity contrast data, dual-energy and photon-counting CT may better guide PINN-based blood flow predictions, potentially improving robustness and accuracy of flow field estimation in clinical  settings\cite{onishi_photon-counting_2024}.

\section{Conclusion}
In this study, we evaluate image- and sinogram-based flow estimation using physics-informed neural networks. 
By evaluating the impact of motion artifacts (via varying gantry rotation speed) and imaging noise (via changes to tube current), we evaluated the ability of PINNs to provide accurate flow estimates in settings where the images of the concentration are not ideal.
Our findings suggest that a sinogram-based approach (SinoFlow) provides more robust performance and should be used when evaluating the use of PINNs to estimate flow from CT data.

\newpage     

\section{Tables}

\begin{sidewaystable}[ht]
    \centering
    \caption{
    \textbf{Comparison of Error Metrics for ImageFlow and SinoFlow at 1, 4 and 10 Hz GRS} }
    \begin{tabular}{|c|c|c|c|c|c|c|}
\hline
 \multirow{2}{4em}{\textbf{Metric}} & \multicolumn{2}{|c|}{\textbf{\rule{0pt}{3ex}1 Hz}} & \multicolumn{2}{|c|}{\textbf{4 Hz}} & \multicolumn{2}{|c|}{\textbf{10 Hz}} \\
\cline{2-7}
& \textbf{\rule{0pt}{3ex}ImageFlow}    & \textbf{Sinoflow}  & \textbf{ImageFlow} & \textbf{Sinoflow}  & \textbf{ImageFlow} & \textbf{Sinoflow} \\
\hline
\makecell{\textbf{Conc.} \\ \textbf{RMSE}} & \makecell{0.375 \\(0.33-0.413)} & \makecell{0.159 \\(0.140-0.174)}& \makecell{0.224 \\(0.209-0.259)} & \makecell{0.133 \\(0.121-0.152)}& \makecell{0.189 \\(0.185-0.192)}& \makecell{0.127 \\(0.118-0.136)}
\\
\hline
\makecell{\textbf{Inlet Vel.} \\ \textbf{RMSE (m/s)}} & \makecell{0.145 \\(0.144-0.161)} & \makecell{0.024 \\(0.023-0.026)}& \makecell{0.063 \\(0.057-0.063)} & \makecell{0.017 \\(0.017-0.018)}& \makecell{0.017\\ (0.015-0.019)} & \makecell{0.016 \\(0.015-0.016)}
\\
\hline
\makecell{\textbf{Inlet Vel.} \\ \textbf{Range (m/s)}} & \makecell{0.109 \\(0.108-0.123)} &\makecell{0.021 \\(0.019-0.023)}& \makecell{0.078 \\(0.059-0.102)} & \makecell{0.014 \\(0.01-0.016)}& \makecell{0.01 \\(0.008-0.116)} & \makecell{0.01 \\(0.009-0.013)}
\\
\hline
\makecell{\textbf{Inlet High} \\ \textbf{Vel. Err (m/s) }} & \makecell{0.195 \\(0.189-0.202)} & \makecell{0.014 \\(0.012-0.0165)}& \makecell{0.075 \\(0.06-0.095)} & \makecell{0.01 \\(0.0096-0.012)}& \makecell{0.024 \\(0.019-0.028)} & \makecell{0.01 \\(0.009-0.014)}
\\
\hline
\makecell{\textbf{Inlet Low} \\ \textbf{Vel. Err (m/s)}} & \makecell{0.085 \\(0.074-0.097)} & \makecell{0.034 \\(0.032-0.037)}& \makecell{0.012 \\(0.007-0.013)} & \makecell{0.025 \\(0.024-0.026)}& \makecell{0.01 \\(0.009-0.011)} & \makecell{0.021 \\(0.021-0.023)}
\\
\hline
\makecell{\textbf{Outlet} \\ \textbf{Ratio}} & \makecell{1.002 \\(0.459-1.108)} & \makecell{1.099 \\(1.074-1.129)}& \makecell{1.190 \\
(0.987-1.250)} & \makecell{1.084 \\(1.072-1.089)}& \makecell{1.034 \\(1.021-1.046)} & \makecell{1.094 \\(1.084-1.106)}
\\
\hline
\end{tabular}
    \label{tab:grs_results_1Hz}
    \label{tab:grs_results}
\end{sidewaystable}

\begin{sidewaystable}[ht]
    \centering
    \caption{
    \textbf{Comparison of Error Metrics for ImageFlow and SinoFlow at 4 Hz GRS and CNR of 12 and 60} }
    \begin{tabular}{|c|c|c|c|c|}
\hline
 \multirow{2}{4em}{\textbf{Metric}} & \multicolumn{2}{|c|}{\textbf{\rule{0pt}{3ex}CNR = 12}} & \multicolumn{2}{|c|}{\textbf{CNR = 60}} \\
\cline{2-5}
& \textbf{\rule{0pt}{3ex}ImageFlow}    & \textbf{Sinoflow}   & \textbf{ImageFlow} & \textbf{Sinoflow} \\
\hline
\makecell{Conc. \\ RMSE} & \makecell{0.292 \\(0.277 - 0.303)} & \makecell{0.289 \\(0.28 - 0.3)}& \makecell{0.237 \\(0.218 - 0.261)} & \makecell{0.142 \\(0.134 - 0.153)}
\\
\hline
\makecell{Inlet Vel. \\ RMSE (m/s)} & \makecell{0.096 \\(0.09 - 0.097)} & \makecell{0.048 \\(0.042 - 0.053)}& \makecell{0.064 \\(0.058 - 0.075)} & \makecell{0.02 \\(0.019 - 0.021)}
\\
\hline

\makecell{Inlet Vel. \\ Range (m/s)} & \makecell{0.101 \\(0.089 - 0.108)} & \makecell{0.073 \\(0.06 - 0.08)}& \makecell{0.09 \\(0.07 - 0.13)} & \makecell{0.013 \\(0.012 - 0.016)}
\\
\hline
\makecell{Inlet High \\ Vel. Err (m/s) } & \makecell{0.128 \\(0.123 - 0.134)} & \makecell{0.012 \\(0.008 - 0.016)}& \makecell{0.092 \\(0.074 - 0.127)} & \makecell{0.013 \\(0.01 - 0.015)}
\\
\hline
\makecell{Inlet Low \\ Vel. Err (m/s)} & \makecell{0.027 \\(0.025 - 0.028)} & \makecell{0.078 \\(0.069 - 0.085)}& \makecell{0.009 \\(0.005 - 0.01} & \makecell{0.027 \\(0.026 - 0.027)}
\\
\hline
\makecell{Outlet \\ Ratio } & \makecell{1.002 \\(0.875 - 1.143)} & \makecell{1.038 \\(0.992 - 1.097)}& \makecell{1.118 \\(0.946 - 1.269)} & \makecell{1.084 \\(1.049 - 1.093)}
\\
\hline
\end{tabular}
    \label{tab:noise_results_CNR12}
\end{sidewaystable}

\begin{sidewaystable}[ht]
    \centering
    \caption{
    \textbf{Comparison of Error Metrics for pulse width of 10 and 50 views at 15\% and 75\% duty cycle under 4 Hz GRS and CNR of 60} }
    \begin{tabular}{|c|c|c|c|c|}
\hline
 \multirow{2}{4em}{\textbf{Metric}} & \multicolumn{2}{|c|}{\textbf{\rule{0pt}{3ex}DC = 15\%}} & \multicolumn{2}{|c|}{\textbf{DC = 75\%}} \\
\cline{2-5}
& \textbf{\rule{0pt}{3ex}PW = 50 views}    & \textbf{PW = 10 views}   & \textbf{PW = 50 views} & \textbf{PW = 10 views} \\
\hline
\makecell{Conc. \\ RMSE} & \makecell{0.34 \\(0.339 - 0.341)} & \makecell{0.257 \\(0.255 - 0.259)}& \makecell{0.142 \\(0.14 - 0.144)} & \makecell{0.139 \\(0.136 - 0.141)}
\\
\hline
\makecell{Inlet Vel. \\ RMSE (m/s)} & \makecell{0.172 \\(0.167 - 0.172)} & \makecell{0.046 \\(0.031 - 0.047)}& \makecell{0.022 \\(0.019 - 0.025)} & \makecell{0.022 \\(0.022 - 0.024)}
\\
\hline
\makecell{Inlet Vel. \\ Range (m/s)} & \makecell{0.099 \\(0.089 - 0.1)} & \makecell{0.063 \\(0.042 - 0.071)}& \makecell{0.019 \\(0.017 - 0.019)} & \makecell{0.019 \\(0.016 - 0.021)}
\\
\hline
\makecell{Inlet High \\ Vel. Err (m/s) } & \makecell{0.215 \\(0.21 - 0.281)} & \makecell{0.013 \\(0.008 - 0.017)}& \makecell{0.012 \\(0.012 - 0.015)} & \makecell{0.013 \\(0.009 - 0.013)}
\\
\hline
\makecell{Inlet Low \\ Vel. Err (m/s)} & \makecell{0.119 \\(0.114 - 0.123)} & \makecell{0.072 \\(0.034 - 0.076)}& \makecell{0.03 \\(0.025 - 0.034)} & \makecell{0.031 \\(0.029 - 0.033)}
\\
\hline
\makecell{Outlet \\ Ratio } & \makecell{0.887 \\(0.817 - 0.967)} & \makecell{1.047 \\(0.915 - 1.248)}& \makecell{1.072 \\(1.055 - 1.094)} & \makecell{1.074 \\(1.062 - 1.082)}
\\
\hline
\end{tabular}
    \label{tab:duty_results_15}
\end{sidewaystable}

\clearpage
\section{Figures}

\clearpage

\section*{Appendix}
\addcontentsline{toc}{section}{\numberline{}Appendix}

\subsection{Computational Fluid Dynamics}
\subsubsection{Boundary Conditions}
\label{subsec:appendix_cfd_bc}

The flow within the vessel was simulated using the in-house CFD code 
TUCAN\cite{moriche_aerodynamic_2017}, to solve the Navier-Stokes equations for Newtonian, incompressible flow in a square computational domain of side length $25H$.
This solver uses second-order finite differences on a Cartesian staggered grid for spatial discretization and a three-stage, low-storage, 
semi-implicit Runge-Kutta scheme for temporal integration. 
The vessel bifurcation is embedded in the fluid computational domain by imposing no-slip boundary condition (zero velocity) on the vessel 
walls using the immersed boundary method \cite{uhlmann_immersed_2005}.  A spatial resolution of $\Delta x/H = 1/100$ was selected based on a grid refinement study conducted using a Poiseuille flow in the same 2D geometry (with occlusion included). 

At the edges of the computational domain, we prescribe a Womersley inflow profile for the parent vessela t $x = 0$, utilize convective 
outflow conditions at $x = 25H$, and free-slip at the top and bottom edges ($y = -12.5H$, and $y = 12.5H$). The inflow boundary condition is prescribed using the pulsatile plane-channel Womersley solution, obtained by 
solving flow between two parallel flat walls separated by a distance $H$, as described in \cite{guerrero-hurtado_efficient_2023}.
The non-dimensional velocity components are expressed as:
\begin{equation}
\begin{split}
\tilde{u} & = \frac{Re}{2} \Delta \tilde{P} \left( \frac{1}{4} - \tilde{y}^2 \right) + \\
              & Re \left[ \frac{\tilde{A} Re}{i \alpha^2}
                        \left( \frac{e^{-\lambda H/2} - e^{\lambda H/2}}
                                    {e^{-\lambda H} - e^{\lambda H}}
                               \left(e^{-\lambda H \tilde{y}} + e^{\lambda H \tilde{y}}\right) - 1
                        \right)
                        e^{i \left(\alpha^2/Re\right)\tilde{t}} \right],
\label{eq_ux}
\end{split}
\end{equation}
and $\tilde{v} = 0$,
where $\tilde{y}=y/H$ is the non-dimensional vertical coordinate of the 2D, $\tilde{t}=t u_c/H$ is the non-dimensional time and $u_c$ is the non-dimensional velocity.
The flow parameters, $u_c = 30 \left[cm/s\right]$ and $H = 1.5 \left[cm\right]$, were selected to represent flow through the inferior vena 
cava (IVC). Normal values of blood viscosity ($\nu=3.8 \times 10^{-6} \left[m^2/s\right]$) and heart rate (70 bpm) were chosen, corresponding to an angular frequency $\omega = 7.33 \;\left[rad/s\right]$. These parameters define the Strouhal number,
$St = \omega H /u_c = 0.37$, and the Reynolds number, $Re = u_c H / \nu = 1184$, which are used to calculate
the constants $\lambda = \sqrt{i Re St}$ and $\alpha = \sqrt{Re St} = 20.83$ (Womersley number) used in Eq. \ref{eq_ux}.
Finally, $\Delta \tilde{P}$ and $\tilde{A}$ are the non-dimensional average and pulsating pressure gradients in the streamwise direction ($x$) and were selected to yield a maximum streamwise velocity, $\tilde u = u_c$.

%

%
%
%
%
%
%
%
%
The pulsatile flow was initialized from Poiseuille flow and run with a Courant–Friedrichs–Lewy number $<0.3$, until the velocity 
field was converged.  Convergence was defined using the maximum velocity field differences between cycles that after 10 cycles were smaller than $5\%$. 

\subsubsection{Concentration Field Solution}
\label{subsec:appendix_cfd_concentration}
To integrate this advection partial differential equation we used a third-order weighted essentially non-oscillatory (WENO) scheme \cite{jiang_efficient_1996} to avoid spurious oscillations as in previous works \cite{gonzalo_span_2022,duran_pulmonary_2023, garcia-villalba_demonstration_2021}.
We integrated the equation in time using the same numerical scheme used to solve the flow.
Since flow and passive scalar problems are decoupled, we used the velocity field from the converged cycle to run several $C_{ca}$ simulations modifying the frequency of the $C_{ca}$ inlet BC
\begin{equation*}
C_{ca}^{inlet} \left( t \right) = \sin ^2 \left( \beta\, St\, \pi\, t \right),
\label{cca_inlet_bc}
\end{equation*}
where $\beta$ is the constant modified to tune the frequency of the contrast agent bolus with respect to the frequency of the Womersley flow.  In our simulations, we considered a case with two bolus per cycle ($\beta = 2$).
\subsection{PINN Loss functions}
\label{subsec:appendix_pinn_losses}

Specifically
$$
\mathcal{L}_{physics} = e^2_1 + e^2_2 + e^2_3 + e^2_4
$$
\begin{align*}
e_1 &= c_t + (u * c_x + v * c_y) 
\\
e_2 &= u_t + (u * u_x + v * u_y) + p_x - \frac{1}{\text{Re}} (u_{xx} + u_{yy})\\
e_3 &=  v_t + (u * v_x + v * v_y) + p_y -  \frac{1}{\text{Re}} (v_{xx} + v_{yy})\\
e_4 &= u_x + v_y
\end{align*}

In the flow equations, \texttt{Re} is the Reynolds number, which indicates the ratio of inertial forces vs. viscous forces. In this work, we used a fixed value $Re=1184$ to match our ground-truth CFD simulations, although  this parameter is also trainable. 
This Reynolds number is representative of mid-size arteries like the IVCs. 
The terms $e_1$ to $e_4$ correspond to the Navier-Stokes equations. 
Specifically, $e_1$ represents conservation of mass, $e_2$ and $e_3$ represent conservation of momentum and $e_4$ represents incompressibility of fluid.

\section*{Supplemental Figures}
\addcontentsline{toc}{section}{\numberline{}Supplemental Figures}

\setcounter{figure}{0}
\renewcommand{\thefigure}{S\arabic{figure}}

\subsection{Impact of Gantry Rotation Speed on Outlet Flow Estimation}

\begin{supfigure}[ht]
    \centering
  \includegraphics[width=\linewidth, page=1]{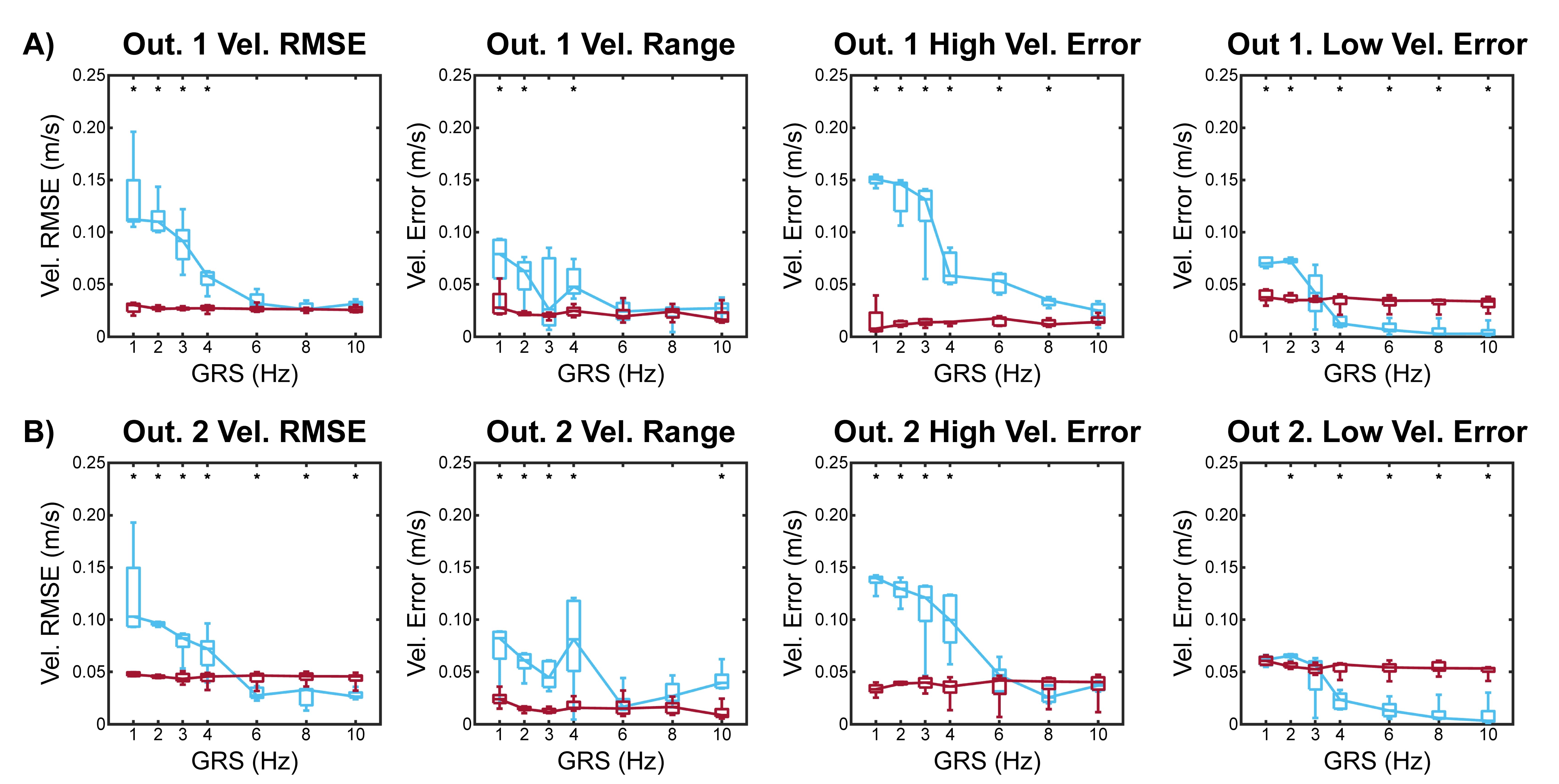}
    \caption{\textbf{Errors in PINN-predicted flow at outlets for different gantry rotation speed.}
    ImageFlow (blue) and SinoFlow (red) predicted flow errors at the two outlets, using boxplots (each box contains 6 data points from the 6 tested gantry starting angle).
    For statistical analysis, black asterisks indicates significant difference found between ImageFlow and SinoFlow results 
    }
    \label{suppfig_GRS_Outlet}
\end{supfigure}

\newpage
\subsection{Impact of Imaging Noise on Outlet Flow Estimation}

\begin{supfigure}[ht]
    \centering
    \includegraphics[width=\linewidth, page=1]{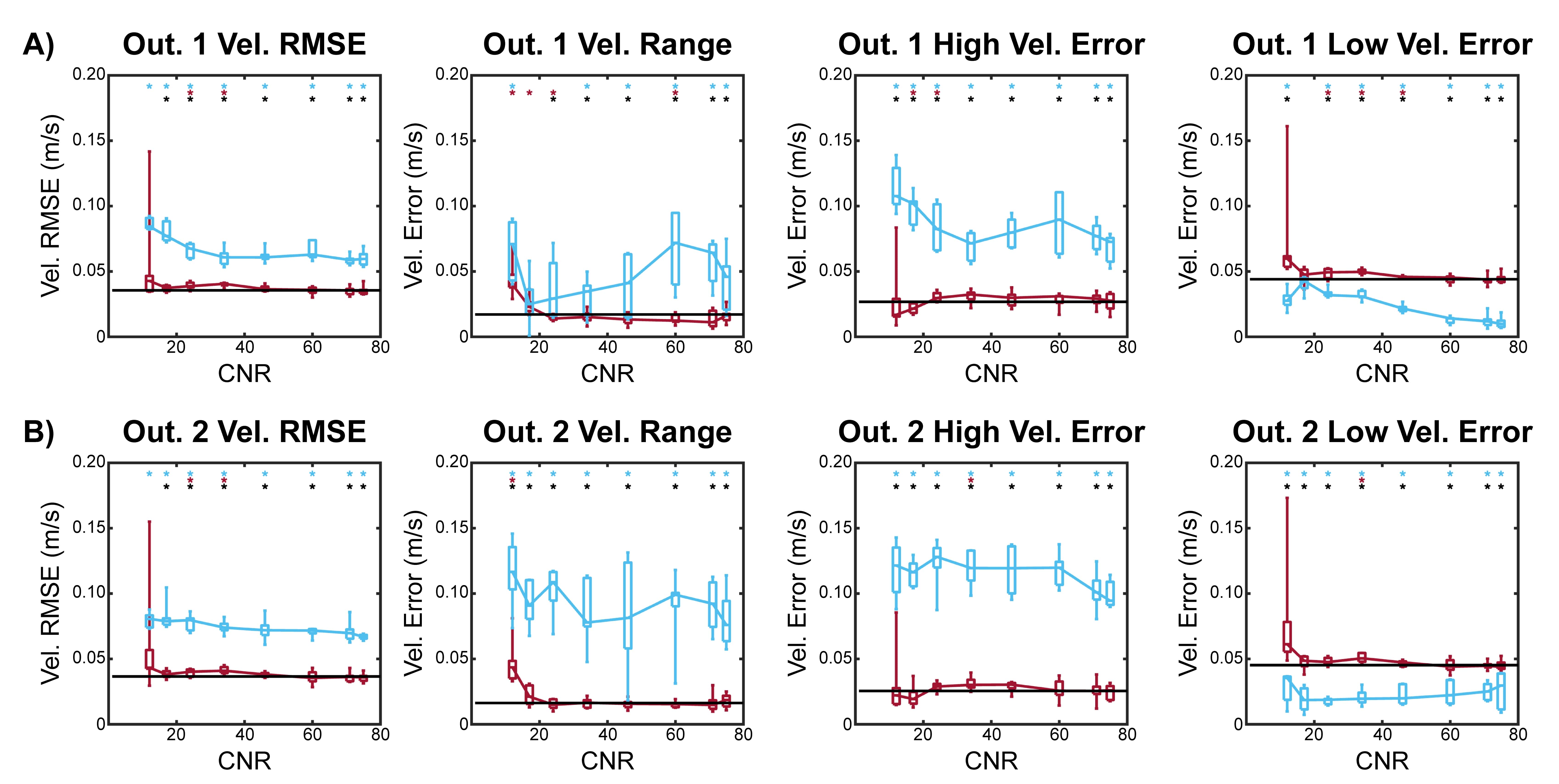}
    \caption{\textbf{Errors in PINN-predicted flow at two outlets at 4 Hz GRS and varying CNR.} 
    ImageFlow are in blue and SinoFlow results are in red. 
    Black lines for the four velocity error panels are the mean noise-free SinoFlow results at 4 Hz. 
    For statistical analysis, black asterisks indicates significant difference found between ImageFlow and SinoFlow results. Blue asterisks (ImageFlow) and red asterisk (SinoFlow) compare different CNR results to the noise-free SinoFlow at 4 Hz (black line).
    }
    \label{suppfig_Noise_Outlet}
\end{supfigure}

\newpage
\subsection{Impact of PM Imaging Parameters on Outlet Flow Estimation}

\begin{supfigure}[ht]
    \centering
    \includegraphics[width=\linewidth, page=1]{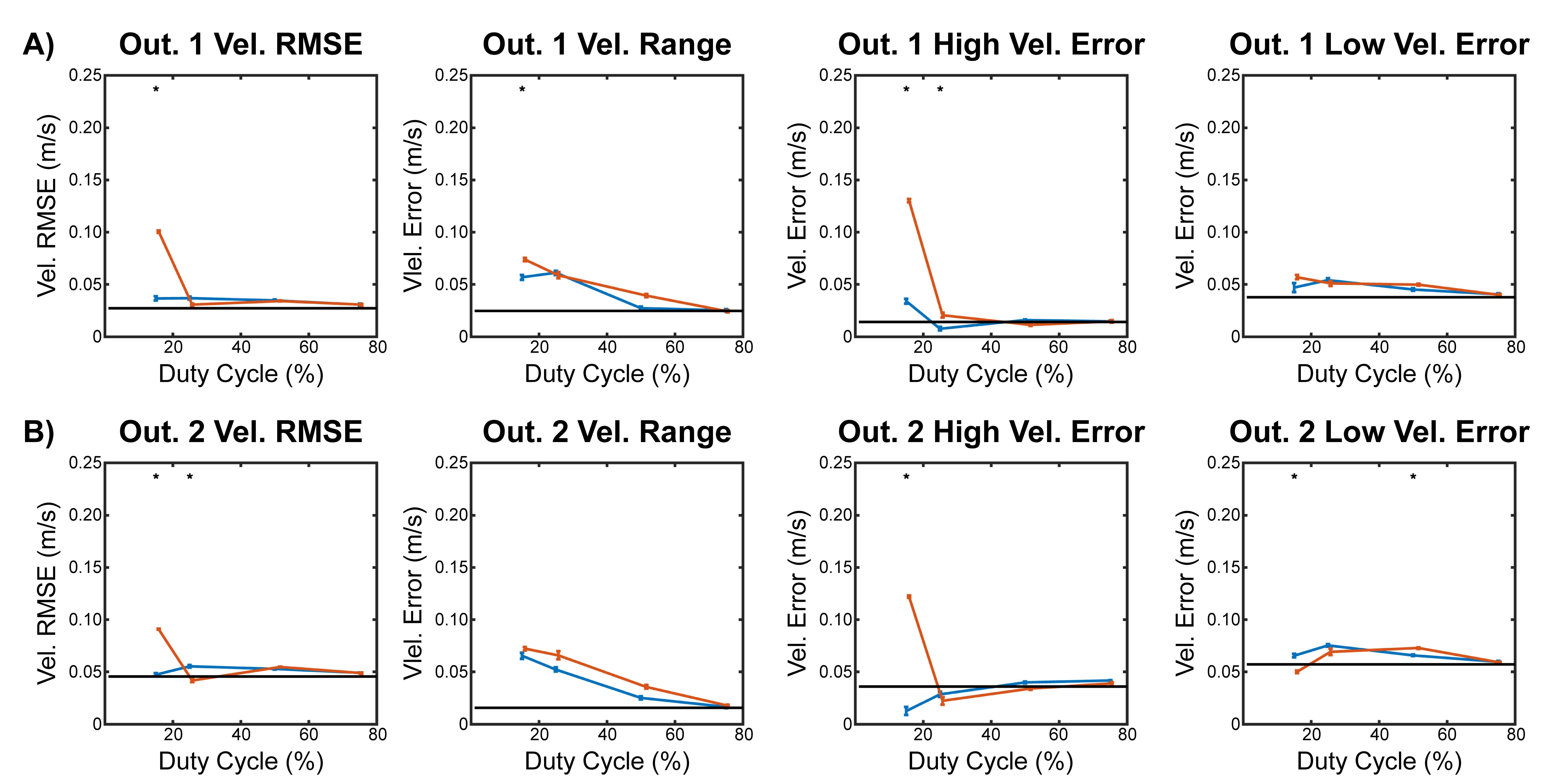}
    \caption{\textbf{Errors in PINN-predicted flow at two outlets for 15\%, 25\%, 50\% and 75\% duty cycle with 10-view (blue) and 50-view (orange) pulse widths.} 
    Each point comprises 6 values (for different gantry rotation positions).
    Black lines correspond to the mean SinoFlow error at GRS=4 Hz GRS and CNR=60.
    Black asterisks indicates significant difference found between between 10-view and 50-view pulse width results.
    }
    \label{suppfig_DutyCycle_Outlet}
\end{supfigure}

\section*{References}
\addcontentsline{toc}{section}{\numberline{}}
\printbibliography[heading=none]










\end{document}